\documentclass[a4paper]{article}
\usepackage[T1]{fontenc}
\usepackage[utf8]{inputenc}
\usepackage{amsthm,amsmath,amsfonts,amssymb}
\usepackage{pdfsync}
\usepackage{enumerate}
\usepackage{graphicx}
\usepackage{epsfig} 
\usepackage{calc}
\usepackage{textcomp}
\usepackage{tabularx}
\usepackage{txfonts}
\usepackage{hyperref}
\usepackage{physics}
\usepackage{color}
\usepackage[authordate]{biblatex-chicago}
\usepackage{authblk}

\title{Black Boxes in Black Hole Imaging}
\author{Juliusz Doboszewski$^{1,2,3}$ and Jamee Elder$^{4,2}$$^\ast$$^\dagger$}

\affil{$^{1}$Institute of Philosophy, Jagiellonian University, Kraków, Poland $^{2}$Black Hole Initiative, Harvard University, Cambridge, USA. $^{3}$Department of Philosophy, University of Bristol, UK. $^{4}$Philosophy Department, Tufts University, Medford, USA.\\

$^\ast$Corresponding author: Juliusz Doboszewski; \href{mailto:jdoboszewski@gmail.com}{jdoboszewski@gmail.com}\\
      
$^\dagger$Authors contributed equally.
}

\begin{document}

\maketitle

\begin{abstract}
We investigate the epistemic opacity of computer simulations and machine learning methods in the context of black hole imaging. We argue that there are forms of opacity---including opacity resulting from the use of machine learning---which do not need to affect the reliability of an inference when it is seen as a part of a broader inferential framework. We propose conditions under which that can plausibly be the case, and discuss how opaque methods can be useful in the context of the (next generation) Event Horizon Telescope. However, we also argue that at least one problematic form of opacity is currently present in black hole imaging: GRMHD models of Sagittarius A* are opaque. This form of opacity signals the limitations of current understanding of the models of this source, and constrains the potential uses of ML models in future observations.
\end{abstract}

\tableofcontents

\section{Introduction}

Ongoing efforts in black hole imaging---primarily by the Event Horizon Telescope (EHT) Collaboration---have been producing high resolution images of black holes with their surrounding accretion disks. These include the first direct image of a black hole shadow\footnote{See \textcite{Skulberg2021} for historical analysis of the EHT and its production of the first image of M87*. See also \textcite{Skulberg-Elder-direct-2025} for historical and philosophical discussion of the use of language of `directness' in black hole research, especially black hole imaging.}, based on radio observations of the center of the galaxy M87 (hereafter, M87*) \autocite{2019EHT_M87_paper1,2019EHT_M87_paper2,2019EHT_M87_paper3,2019EHT_M87_paper4,2019EHT_M87_paper5,2019EHT_M87_paper6}; polarization of M87* emission \autocite{2021EHT_M87_paper7,2021EHT_M87_paper8}; further images of M87* using data from a subsequent observing campaign \autocite{2024EHT_M87_persistent}; as well as images of the supermassive black hole in the center of our own galaxy, referred to as Sagittarius A* (hereafter, Sgr A*) \autocite{EHT2022_SgrA_paper1,EHT2022_SgrA_paper2,EHT2022_SgrA_paper3,EHT2022_SgrA_paper4,EHT2022_SgrA_paper5,EHT2022_SgrA_paper6}, and its polarization \autocite{EHT2024_SgrA_paper7polarization,EHT2024_SgrA_paper8polarization-interpretation}.

The observations underlying these images are short wavelength radio waves collected by radio telescopes around the world and combined using a technique known as Very Long Baseline Interferometry (VLBI). This means that the telescopes, separated by large distances, are coordinated in a virtual array whose resolution (in the EHT case) approximates that of a single telescope with the diameter of the Earth. Each pair of elements in the array provides a sampling point, which for a sparse array implies a sparse coverage of the source. Moreover, the data are noisy and have to be subjected to extensive calibration and reduction procedures. These features make the reconstruction of an image of the source emitting radio waves a highly non-trivial and, in general, ill-posed problem: sparse information provides weak constraints, and as a result, a large number of very different images and models of the source are compatible with such data.

In order to recover a unique and informative image and make further inferences from the data, auxiliary assumptions about the image (for example, concerning its overall smoothness or shape) need to be built into the imaging process. Consequently, any bias introduced by these assumptions also needs to be controlled. This is crucial for the results to be able to discriminate between competing hypotheses about the source, as opposed to merely confirming modelers' biases.\footnote{For a recent philosophical take on these observations, see \textcite{Elder-Doboszewski2024}. In this paper we argue that robustness---understood as convergent results across of independent lines of analysis---is employed in minimizing the bias introduced by theoretical assumptions in the M87* imaging process. \textcite{Elder-Doboszewski2025} also provide a related discussion of the ways in which parameter estimation processes employed by the EHT collaboration are theory-laden.} Imaging algorithms, which implement different approaches to solving the problem of recovering an image from the data, are one of the points where such assumptions enter into the process. These algorithms are used to take sparse samples of visibility data (the Fourier transform of the source emission) and `solve' the ill-posed problem to produce an image. Such algorithms are sometimes even seen by practitioners as part of the telescope, since they take the sparse sampling and produce an image \textit{as if} based on a single telescope spanning the Earth. Indeed, without imaging algorithms the usefulness of data collected by the telescope would be at best highly limited. Further algorithms are used to model the source and estimate its parameters (such as its mass, spin, or type of accretion) on the basis of the images.

Recent work in machine learning (ML) has raised the possibility of using algorithms that implement ML methods to play such roles. The most important motivation for doing so is solving (or at least mitigating) a problem faced by current imaging methods: data analysis bottlenecks. The problem, in short, is that data are being collected faster than they can be analyzed, leading to a growing backlog of data for which results have not been published. 

VLBI observations of black holes face a data analysis bottleneck due to the abundance of data they collect, combined with the difficulties of analyzing it. Consider the following numbers. The EHT array observes a number of targets for \(\sim \)week/year (including successful runs in 2017, 2018, 2021, 2022, 2024, and in 2025). Possible further extensions of the EHT array\footnote{Such as the Black Hole Explorer with an orbiting telescope component \autocite{johnson2024bhex-motivation}, or a less sparse next generation EHT array. Throughout this paper we use the phrase ``black hole imaging'' as subsuming observations with these possible future instruments. Inference methods are similar for these variations of the EHT array, and concerns about scalability, data volumes, bias, and opacity arise for all of them.} will provide significantly more observation time. For instance ``[i]n Phase 1, the ngEHT will add three or more dedicated telescopes to the current EHT, with primarily dual-band (230/345 GHz) observations over \(\sim \)3 months per year. In Phase 2, the ngEHT will add five or more additional dedicated telescopes, with simultaneous tri-band capabilities (86/230/345 GHz) at most sites and observations available year-round.'' \autocite[3]{ngeht-key-science-goals}. However, data analysis is already lagging behind the EHT data, and will face even greater challenges keeping up with expanded data collection. The famous 2019 M87* image is based on 2017 data; the 2024 M87* image on 2018 data \autocite{2024EHT_M87_persistent}; and the 2025 on 2021 data \autocite{EHT-2025-M87}. The only image of the second main target, Sgr A*, is from the 2017 observations. These are merely a few data points, but the trend already looks dismal. Since data processing is already a bottleneck, developing reliable imaging and inference algorithms that decrease processing time will be crucial for scaling up current VLBI observations of black holes. 

This bottleneck problem presents a strong motivation for exploring ML approaches to black hole imaging. After all, if there's one thing ML is truly good for, it is quickly analyzing vast amounts of data.\footnote{Other methods for speeding up the analyses are also being developed, for example relying on high-speed programming language Julia.}

However, a burgeoning literature in the epistemology of ML highlights how these methods can have associated risks of opacity (due to complexity of the model, its training process, and representational roles) and built-in bias (due to large number of assumptions and/or composition of the training dataset). These worries about ML extend concerns about dealing with opacity and bias in complex computer simulations. Both of these types of models might be seen as black boxes, whose inner workings are difficult to see through and control. Increasing adoption of ML in scientific inference in general and astronomy in particular further underscores the importance of understanding the potential applications and limitations of such methods.

Given this, we think it is important to ask: In what ways do problems of opacity and bias already occur in black hole imaging? To what extent could the proposed uses of ML methods impose additional epistemological problems? And, ultimately, how should these methods be implemented in the context of black hole imaging (if at all)?

Our answers to these questions are (roughly stated) as follows. 

First, some forms of opacity can be identified in the use of General-Relativistic Magnetohydrodynamics (GRMHD) simulations (of hot plasma in a strong gravitational field) to make inferences about the properties of highly variable sources. The EHT's current imaging procedures are not implicated, but modeling of some sources is. Bias is also arguably well-controlled for current imaging pipelines.

Second, some applications of ML to imaging \textit{do} present additional problems, although not for the reasons one might initially think; the problems derive from the upstream opacity and bias of GRMHD simulations, rather than from the opacity of ML methods.

Third, our contextual approach to ML methods implies that while they could be among the imaging algorithms employed by astronomers, they should only be trusted in epistemic contexts where other methods provide independent checks on results. In such contexts, there is little need to push for the development of explainable/interpretable methods.

In what follows, we first, in section \ref{sec:opacity}, provide an overview of philosophical concerns about opacity and bias in both complex computer simulations and ML methods. In section \ref{sec:sgra} we discuss how the opacity of GRMHD simulations manifests itself in inferences about Sgr A*, the supermassive black hole at the center of our own galaxy. In section \ref{sec:ML_astro} we discuss some of the proposed implementations of ML methods in large astronomical catalogs and in black hole imaging, drawing out the benefits and potential problems with these methods. Then, in section \ref{sec:ourtake} we provide a proposal for how to implement opaque ML to draw secure inferences about supermassive black holes. In section \ref{sec:conc} we briefly conclude with some reflections on the widespread push for explainable or interpretable\footnote{This distinction is clarified in section \ref{sec:opacity}.} ML methods. In particular, the case of black hole imaging supports a view according to which the reliability of opaque methods can be established in the absence of explainability or interpretability, assuming that the broader epistemic context satisfies certain conditions.

Overall, we provide criteria for the reliable use of ML in black hole imaging based on an analysis of the broader epistemic context in which that imaging takes place. In doing so, we extract from philosophical literature on computer simulations and machine learning a normative framework of possible use to astronomers considering taking up new ML tools.\footnote{It is worth noting that the next generation EHT presents a unique opportunity for philosophers, as well as other scholars in the humanities and social sciences, to engage with the inference methods that are prioritized by scientists. This is because such scholars are embedded in the project from the get-go as one of the `Science Working Groups' \autocite{ngeht-hpc-2023}. This opens up potential avenues in which philosophical argumentation might influence the collaboration's activities. With this in mind, conclusions from the philosophical literature on ML may be communicated to interested parties in the larger collaboration and put to practical use. Accordingly, in this paper we take up the dual aims of (1) using the case of black hole imaging to inform the philosophical literature on ML methods in science, and (2) taking on lessons from this literature to develop philosophically motivated prescriptions that might feed directly into scientific practice.}

For philosophers of science, we think that a key takeaway is that the reliability of opaque methods can be established even in highly challenging epistemic circumstances. Black hole astrophysics lacks direct access to its targets (due to extreme distances); probes extreme physical regimes (that are otherwise unobservable); and relies heavily on computationally expensive (and sometimes opaque) simulations for generating theoretical accounts of these systems. Yet even in such cases, the reliability of opaque methods can be established. This supports a more general philosophical view.  If --- as we argue --- one can establish the reliability of opaque machine learning algorithms even in this hard case, then, plausibly, one could do that in many other, more down-to-Earth contexts. In other words, we think that this case provides grounds for a form of optimism about the reliability of opaque methods. Assuming that the appropriate contextual factors (such as those we discuss in section \ref{sec:ourtake}) hold, a good handle on the reliability can be had without opening the black box. In this sense, we see our analysis of the black hole imaging as continuous with the work of \textcite{Duran-Formanek,Duran-Jongsma2021,sullivan2020understanding,meskhidze2023can,martinking-MI,koberinski-ml-paper}.

\section{Epistemic Opacity in Simulations and Machine Learning}\label{sec:opacity}

`Epistemic opacity'---or simply `opacity'---is a focal notion in philosophical discussions of computer simulations and machine learning. Opacity is often taken to present a challenge for establishing trust in, and gaining understanding from, the results obtained with the opaque model. This section reviews philosophical literature related to these concerns in order to pose the question: should one be concerned about the prospect of using opaque methods in astrophysical inferences? In other words, should one insist on explainable or interpretable models in astrophysical contexts?

The idea of epistemic opacity was first introduced to the philosophy literature by Paul Humphreys' discussion of computer simulations:  
\begin{quote}
``In many computer simulations, the dynamic relationship between the initial and final states of the core simulation [...] is epistemically opaque because most steps in the process are not open to direct inspection and verification'' \autocite[147-8]{Humphreys2004}.\footnote{See, for example, \textcite{beisbart2022philosophy} for discussion of Humphreys' development of the idea.} 
\end{quote}
In these situations, the opacity arises due to the complexity of the computation that generates the simulation; the number of epistemically relevant computational steps exceeds what a human can follow. 

In the context of climate modeling, \textcite[170]{Winsberg2018} has argued that:
\begin{quote}``climate models are, to some degree epistemically opaque to their users and developers. That is: the features that account for the successes and failures of climate models in simulating certain aspects of the climate system are not always fully understood and appreciated by modelers who use and make them.''
\end{quote}
This characterization of simulation opacity is more general in the sense that it concerns modeler's uncertainty about why a simulation works---not necessarily (only) due to the number of computational steps. We will use this characterization in our argument establishing the opacity of GRMHD simulations of Sgr A* in section \ref{sec:sgra}.

The problem of opacity extends to many approaches to ML as well, and in particular to deep neural networks (DNN) with their complex network architecture and learning process. For example, \textcite[4]{beisbart2022philosophy} argue that ``the computations done during the training phase as well as those that lead from a specific input to an output in deployment cannot be followed by humans within reasonable time''.\footnote{Additionally, there is some debate about which features of DNNs are sufficient for them to be opaque (and in what sense, relative to distinctions made by \textcite{Creel2020}, \textcite{Boge2022} and others). \textcite{Søgaard2023} provides a detailed evaluation of how five features of DNNs---size, continuity, nonlinearity, instrumentality, and incrementality---do and do not lead to different forms of opacity. In addition, in the context of uses of ML in cosmology, \textcite{meskhidze2023can} makes a different kind of distinction: between black boxes as objects exhibiting opacity from the ``method of ignoration', or black boxing as a methodology---which involves deliberately treating a model like a black box in order to ignore details of the model that are thought not to matter for a given purpose.}

\textcite{Creel2020} has recently broadened the discussion of opacity and transparency by distinguishing between three different types (or sources of) of transparency based on our knowledge of (1) how the algorithm functions as a whole, (2) how the algorithm is realized in code, and (3) how the program was run, including the hardware, input data, etc. Creel argues that we need to consider all three when attempting to reduce opacity (we will see an example of that in our discussion of PRIMO in section \ref{sec:ML_BH_imaging}).\footnote{There are also a range of other, related but inequivalent, distinctions on offer here. \textcite{Boge2022} provides a different distinction between two ``dimensions'' of opacity: h-opacity, or opacity concerning \textit{how} a deep neural network learns, and w-opacity, concerning \textit{what} it learns. \textcite{Sullivan2022_inductive_risk} distinguishes between external and internal opacity, where these have to do with external validity and internal validity respectively (the latter has to with checking whether the counterfactuals --- what would model do under some circumstances --- extracted from the model actually capture how the model works). \textcite{Søgaard2023} distinguishes between inference-opacity and training-opacity, where the former concerns why a DNN gives a certain output \textit{y} given input \textit{x}, and the latter concerns why a DNNs parameters take the values they do based on the training data.}

When a model or method is opaque, it is sometimes said to be a ``black box'': although we can see the inputs and outputs, we cannot see why the model produces the outputs that it does. As we have just seen, one of the main reasons that models are epistemically opaque---and thus black boxes---is due to their complexity being beyond our comprehension. However, black boxes can also arise in other ways. There are many possible sources of opacity, arising not from nature of the model, but for others reasons, such as restrictions on access to the model. Opacity can also be perspectival, arising only for some users of an algorithm. In our discussion, we are largely interested in opacity and bias arising from the complexity of the computation (including the training phase of ML) which affects all users equally.\footnote{\label{fn:compas1}\textcite{Burrell2016} identifies three sources of opacity: scale and complexity; technical illiteracy; and corporate or state secrecy. In a similar vein, \textcite{Rudin2019} discusses how models can be black boxes due to complexity (rendering them opaque to everyone) or due to their being proprietary (rendering them black boxes to those without proprietary access). For example, COMPAS (``Correctional Offender Management Profiling for Alternative Sanctions''), an algorithm used by U.S. courts to estimate the likelihood of recidivism, is a black box of the latter type, though its documentation indicates that it is not a black box of the former type \autocite[209]{Rudin2019}.}

Opacity leads to worries about model reliability and limits on scientific understanding.
First, the extent to which we trust a model is often thought to be linked to its transparency. As \textcite[2]{watson2022conceptual} puts it, opaque models raise ``fundamental issues of trust. How can we be sure a model is right when we have no idea why it predicts the values it does?''. In contrast, it is more straightforward to establish trust in a transparent model because we are able to understand how it works and evaluate the conditions under which it will produce reliable inferences. 
Second, the use of ML, and of other opaque models, is often seen as placing limits on the scientific understanding of the physical system under investigation. This is because (in contrast to transparent models) extracting explanations and various counterfactuals from opaque models is difficult, as opacity limits the possibility of obtaining answers as to why the model returned the particular outcome. However, both of these points are controversial.

First, trust in a method or process may be established without understanding how it works. This is a familiar point in the philosophy of experiment when it comes to scientific instruments (consider, e.g., \textcite[ch.11]{Hacking1983}'s discussion of microscopes). With respect to computations (algorithms, simulations, etc.) \textit{computational reliabilism}, a version or application of process reliabilism, provides an account of the reliability of algorithms that depends only on their propensity to produce trustworthy outputs: ``[i]n a nutshell, [computational reliabilism] states that researchers are justified in believing the results of artificial intelligence (AI) systems because there is a reliable process (i.e., the algorithm) that yields, most of the time, trustworthy results'' \autocite[332]{Duran-Jongsma2021}.\footnote{See \textcite{Duran-Formanek} for an elaboration of computational reliabilism. We thank Emanuele Ratti for pointing us toward this body of work.} On this view, understanding the inner-workings of a black box is not necessary for establishing its reliability. Embedding a tool---whether a physical instrument, simulation, or ML algorithm---within a broader epistemic context allows us to evaluate the tool based on its performance; if it consistently enough produces the right outputs for a given input, then we can consider it to be reliable under some specified range of conditions. How to go about  appropriately evaluating the performance of opaque ML methods in a given context is then a further separate question (which we tackle in section \ref{sec:ourtake}.) This overall approach resonates with our position on (non-ML) EHT imaging algorithms in \textcite{Elder-Doboszewski2024}.

Second, the supposed trade-off between opacity (or black boxes) and understanding is controversial. \textcite{meskhidze2023can} argues that when cosmologists use ML alongside N-body cosmological simulations, the ML models do confer scientific understanding. In part, this is possible because the N-body simulations already involve structural modeling, incorporating background knowledge about the physical features that are important for large-scale structure formation in the universe. The understanding gained from ML is grounded in this existing (non-opaque) modeling. This, Meskhidze argues, demonstrates the need to consider the broader methodology in which ML methods are operating when deciding whether ML provides scientific understanding. 
In a similar spirit, \textcite[124]{sullivan2020understanding} argues that it is not complexity or opacity \textit{per se} that limits scientific understanding, but rather what she calls ``link uncertainty''. Link uncertainty is characterized as ``lack of scientific and empirical evidence supporting the link connecting the model to the target phenomenon''.\footnote{In other words, link uncertainty concerns difficulties in establishing the external validity of the opaque model.} 
Sullivan uses a framing relying on the notion of representation, which in our view is not entirely satisfactory for something like imaging algorithms, which might be better seen as inferential tools than as representations of their targets (see \textcite{Elder-Doboszewski2024} for a discussion of this view).
Nevertheless, in our assessment of opacity, we take ourselves to be building on work of the kind done by, e.g., \textcite{Duran-Jongsma2021,sullivan2020understanding,meskhidze2023can} and others in paying attention to the broader scientific methodology being implemented around a ML algorithm and the extent to which this mitigates or exacerbates concerns about black box models.

The problems associated with epistemic opacity have motivated a variety of programs aimed at making processes behind ML predictions intelligible to humans. Some such programs---so-called ``explainable AI''\footnote{See, e.g., \textcite{molnar2020interpretable} for an introduction to a large variety of approaches to producing explainable or interpretable ML.}---are aimed at a figurative prying into the black box to extract some kind of explanation about what it is that a ML model is doing. Explainable AI often involves post-hoc attempts to model the black box model in a way that mimics its results and is intelligible to humans. 
However, explainable AI faces some important conceptual and practical problems. \textcite{Rudin2019} argues that ``explainability'' is a misleading terminology; what the new model really seems to be is a how-possibly explanation that replicates (well enough) the black box model's results---it tells us one way that the black box \textit{might} work, rather than how it actually \textit{does} work. One should therefore be suspicious of any claim that a new ``explanation'' model really offers an explanation of the original black box model's outputs.\footnote{\label{fn:compas2}The ProPublica critique of COMPAS \autocite{ProPublica_Machine_Bias} was criticised on these grounds as failing to actually show what they claimed---that COMPAS is biased against black people---because although their explanation model replicates COMPAS results and does exhibit such bias, they cannot show that their model actually reflects how COMPAS makes decisions \autocite[208]{Rudin2019}.} 
In addition, \textcite{watson2022conceptual} argues that current approaches to explainable AI have three problematic features: ambiguity with respect to their target, lack of attention to severe testing, and treating explanations as static derivables rather than as dynamical interactions between various parties.\footnote{\textcite{watson2022conceptual} uses the general terminology of ``interpretable'' AI, which we reserve (following Rudin) for inherently interpretable models. However, since the main arguments of Watson's paper specifically target post-hoc explanations of ``some underlying target function'', his arguments primarily concern (what we are here calling) ``explainable'' AI. Note, however, that this use of ``interpretable'' does not clearly align with the physics literature use of that term in the discussion of R2D2 in section \ref{sec:ML_BH_imaging}.} 

Given these issues with explainable AI, \textcite{Rudin2019} argues that (at least in many contexts) concerns about opacity should lead us to eschew black box models altogether in favor of inherently ``interpretable'' models---models that are transparent and intelligible from the beginning. However, setting up interpretable models with accuracy equivalent to their black box counterparts requires significant subject-matter expertise to be built in. This means that there is a trade-off between interpretability and efficiency, understood in terms of the specialist work needed to set up an interpretable model.

However, Rudin argues against a generic trade-off between interpretability and computational performance; with some additional work, it is generally possible to develop interpretable models with comparable accuracy to opaque models. This performance/interpretability trade-off is somewhat controversial in physics as well. In the LHC context \textcite[175]{guest2018deep} say that (when a neural network is involved) ``generically we should anticipate a trade-off between performance and interpretability''. However, \textcite[1]{dold2022evaluating} find that their ``evaluated machine learning models are capable of producing catalogs of a similar quality as the existing ones which were constructed from mixture modeling and structural fitting''; so it seems that at least in some cases the trade-offs do not need to arise. This also seems to be true in some medical applications: see e.g., \textcite{shanklin2022ethical}.

The question for our purposes here is thus the following: ought we (either in general, or at least in some particular cases, such as astrophysical contexts and ngEHT in particular) favor or even demand either explainable or interpretable ML? Or, could we sometimes be satisfied with the efficiency and accuracy of black boxes? More generally, how should lessons from philosophical discussions of opacity in ML be applied to the case of VLBI-based black hole research? And could there be any methodological lessons that might be drawn in the opposite direction?

With regard to at least some of the intensely-debated philosophical questions concerning ML, astronomy can be seen as a somewhat easier case than many areas of science. This is because astronomy is a socially low-stakes setting when compared to e.g., the use of ML in healthcare, education, or the criminal justice system.\footnote{On ethical issues related to ML, see for example, \textcite{Vaassen2022,Goetze2022,Vredenburgh2022,Baum-etal2022}. 
} In these areas there are clear ethical dimensions implicated in discussions of, say, algorithmic bias or explainability/interpretability. In the COMPAS example (see footnotes \ref{fn:compas1}, \ref{fn:compas2}), there are good reasons to prefer interpretable models: opacity makes it difficult to determine whether a particular sentencing or probation decision is fair; and individuals may have a right to know why they received a given decision (in particular, they may deserve assurance that it was not based on their racial identity). Although astronomy also has to grapple with a range of ethical issues---most prominently, those based on its history rooted in colonialism---such issues are not of immediate relevance for the applications of ML to black hole research.\footnote{For discussion of these ethical issues in physics and astronomy, with special attention to the case of building new telescopes on Maunakea in Hawai'i, see e.g., \textcite{swanner2017instruments,kahanamoku2020native,Prescod-Weinstein2021,ruland2024sustaining,dill2025fork}. Some discussion of issues related to ethical telescope siting in the case of the ngEHT specifically can be found in \textcite{Marcoci-etal2023} and \textcite[section 2.8.1]{ngEHT_KSG}.} So even if ML methods would introduce bias into scientists' inferences about some naturally occurring phenomena, such bias is unlikely to cause anyone serious harm. 

Although the ethical implications of using ML may be minimal in astronomy, the epistemic considerations are a different matter. In general, astronomy faces significant epistemic challenges: astronomers are often dealing with sparse data over small populations, small control groups, and no control over interventions. Therefore mitigating modelers' bias and ensuring the reliability of inference methods are both crucial in establishing trust in the outcome of analyses. Moreover, as already mentioned above, in black hole imaging with VLBI specifically the imaging algorithms are indispensable for getting useful results. If these were opaque, then perhaps the reliability of the inference tool---and, in a way, of the telescope as a whole---could be questioned, on the grounds that extracting the answer as to why the algorithm returned this particular outcome might be impossible due to the algorithm's opacity.

The key question then is which methods should be developed and implemented in the context of black hole imaging. Should one require that ML for the EHT satisfies some explainability benchmarks, or simply interpretability? If so, from the get-go, or at some later stage (perhaps using ML on new real data)? We should note here that in the EHT software ecosystem such a decision is an abstraction. In this context, some imaging algorithms and parameter extraction methods are already an industry standard (e.g., CLEAN); others have already been applied to real EHT data (e.g., eht-imaging); and numerous others have been developed by various groups and individuals, and tested on both synthetic and real data (e.g., through imaging challenges and analysis of publicly-available real datasets). The CHIRP algorithm, developed by \textcite{Bouman-etal2016_CHIRP} is a prominent example of an alternative imaging approach that arguably outperforms CLEAN, but was not used to produce the 2019 image. The complete explanation of why the EHT or future ngEHT adopts the particular inference methods includes a combination of epistemic and social factors, the full analysis of which is beyond the scope of this paper (and would likely encompass significant historical and sociological analysis). Nevertheless, treating this question as if there were a single decision point and decision-maker making a choice about whether ML in black hole imaging could be done with black boxes, explainable models, or interpretable models does provide a useful framing for thinking about the conceptual reasons in favor of particular choices. Accordingly, in the remainder of this paper, we will provide analysis aimed at evaluating epistemological and methodological aspects of such a decision.

Our approach is to focus on the broader inferential context in which ML is used when evaluating the extent to which opacity should be considered a problem. This approach resonates with \textcite[113]{sullivan2020understanding}'s views about gaining understanding from a deep neural network (``DNN''): ``when we consider DNN models and how their opacity may prevent understanding of phenomena, we cannot consider the model in isolation'' as well as the position of computational reliabilism \autocite{Duran-Formanek,Duran-Jongsma2021}. We are also influenced by the broader philosophical literature concerning models and simulations that emphasizes the contextual reliability of a model for a particular inference over concerns about representational accuracy (e.g., \textcite{Morgan-Morrison1999,parker2020model}). 
From this vantage point, we will argue that trust in an opaque method can---under the right conditions---be convincingly established. 
Furthermore, the case of black hole imaging demonstrates how reliability can be established even in challenging epistemic circumstances. This provides reason to be optimistic about the availability of similar lines of argument in a wide range of contexts. 

We turn to the application of ML to astrophysics in sections \ref{sec:ML_astro} and \ref{sec:ourtake}. But first, we will argue that reliance on complex computer simulations in black hole imaging already involves a form of opacity (section \ref{sec:sgra}).

\section{Secrets of the Gentle Giant: Epistemic Opacity and Sagittarius A*}\label{sec:sgra}
 
Before we consider possible roles of opaque methods in the future of black hole imaging, let us first consider the current state of the art in VLBI observations of our nearest supermassive black hole, the ``gentle giant'' Sgr A*.\footnote{Sgr A* is called a ``gentle giant'' because of its very low accretion rate, in contrast to the insatiable appetite (higher accretion rate) of M87*.}

This might seem like a detour in our discussion of opacity and ML. However, modeling of Sgr A* through GRMHD simulations displays an actual form of opacity in physics, which is philosophically interesting on its own. Moreover, insofar as GRMHD simulations are used in the training of ML (which is the case), their limitations will be inherited by inference methods downstream. Their opacity provides a further source of uncertainty in ML, and limits their uses in various domains.

The first EHT images of both M87* and Sgr A* were based on data collected in 2017. However, the images of Sgr A* were released much later (in 2022 vs. 2019). This is due, in part, to the challenges of imaging a source that is variable on short timescales (i.e., one that is changing intra-hour, so during the course of the observations). Despite this additional challenge, the imaging process for Sgr A* was, broadly speaking, similar to that of M87*; a number of alternative imaging methods were used, with each used to produce a range of images that were compatible with the data. 

Each imaging method relies on sophisticated imaging algorithms that allow images to be produced from the sparse, noisy data collected by the array. A useful analogy here is with a camera that collects only a small fraction of the light entering its aperture. Imaging algorithms make up for the lost light, allowing the camera to produce an image based on limited information. Taking a photo with a digital camera---such as with an iPhone camera---combines the data collection and data processing when producing an image. In contrast, taking a photo using the EHT array separates these processes. Nonetheless, imaging algorithms can be thought of as part of the ``camera'', and the EHT imaging process, comprising data collection and subsequent imaging pipelines, is the analog of taking a photo.\footnote{Peter Galison (see e.g., \textcite{Galison2021MoMA}) argues that this is not mere analogy; the EHT images are, on his analysis, genuinely photos. Just as iPhone algorithms build in sophisticated data processing to make up for their physical limitations (e.g., small lens), the EHT uses data processing to make up for gaps in the array.}   
Since each imaging algorithm must build in additional information to make up for the gaps, it is important to check that this doesn't introduce artifacts---i.e., that the images are accurate and can be used to make reliable inferences about the black holes they represent. 

The EHT Collaboration establishes the reliability of their images by comparing the results of different imaging methods.
The outputs of these different imaging pipelines are compared and combined to produce a single final image.\footnote{\label{fn:sgr_vs_m87}The details are a bit different for Sgr A* compared to M87* in two chief respects: (1) Sgr A* included an additional imaging pipeline, based on a method of Bayesian posterior sampling using the THEMIS modeling framework, in addition to the DIFMAP algorithm (CLEAN method) and the eht-imaging and SMILI algorithms (Regularized Maximum Likelihood method) used for M87*; and (2) each of the methods was used to produce a population of reconstructed images. The images from this overall population were then grouped into ``clusters'' exhibiting similar source morphology---across methods. This included 3 clusters exhibiting a ring and 1 cluster that did not. Overall, the results strongly favor a ring.} For both Sgr A* and M87*, image properties are considered to be reliable to the extent that they are robust across different imaging methods (and variation in parameter choices within a given method). From a philosophical standpoint, \textcite{Elder-Doboszewski2024} discuss the robustness argument for confidence in the M87* image in depth. While the details differ (see footnote \ref{fn:sgr_vs_m87}), the broad strokes of the argument are the same for Sgr A*. 

The upshot is the following: while the results exhibit some differences across imaging methods and parameter choices, the image of a ring is preferred. Some features of the images (such as the positions of bright hotspots on the ring) lack the robustness exhibited by the ring feature; they tend to vary, depending on the method used to make the image and the specific parameter choices made during that process. Such features are treated as artifacts of the imaging process and are not trusted.

None of the imaging pipelines build in strong theoretical assumptions about the source system. In particular, the imaging process does not rely on results from GRMHD simulations, the standard simulations for modeling accretion around a supermassive black hole. This means that any epistemic concerns about these simulations do not carry over to the first Sgr A* images.  
However, the physical interpretation of the images (e.g., inferences about the mass of the supermassive black hole) does implicate GRMHD simulations, so any inadequacies of these simulations may bias inferences about the source system. As it turns out, there are reasons to be cautious about the empirical adequacy of GRMHD simulations for Sgr A*.

In what follows, we focus on tests of astrophysical models of Sgr A*. In contrast to images (produced by various imaging pipelines), we use the term ``models'' of Sgr A* to refer to models ascribing physical properties to the source system. For example, a particular GRMHD simulation models the source system as having a set of specific physical properties by building in physical assumptions about the black hole (e.g., its mass and spin) and the accretion dynamics.\footnote{GRMHD simulations can also be used to produce images. Roughly, this involves simulating the evolution of the source and tracing the rays of light connecting the physical source to the observer. One can then take snapshots of the resulting movie to get an image. While this is a kind of black hole imaging, we reserve the term ``black hole imaging'' for images produced from observational data using VLBI.}

To test the adequacy of models of Sgr A*, the EHT Collaboration considers whether the best GRMHD models of this system are consistent with observations.
In their analysis, \textcite{EHT2022_SgrA_paper5} identify eleven observational constraints---roughly, facts drawn from observations that a model needs to get right in its representation of Sgr A*. These constraints are thought to be uncorrelated with each other, can be simulated with the models of that source, are based on emission produced near the main source of 230 GHz emissions, and are measured around the same time as the EHT 2017 observations. In short, they are independent of one another, relevant to our physical understanding of the source, and amenable to comparison with GRMHD models. 
However, they are highly heterogeneous, involving different observations, systematic errors, assumptions about the underlying astrophysics of the system, and physical parameters describing ways the system might be.   

The constraints are split into three classes. 
The first of these consists of five constraints derived from the EHT data (source's size, morphology of the visibility amplitude, and three parameters from the best-fitting ring image model). 
The second consists of four constraints derived from other observations of Sgr A* (flux density and source's size at 86 GHz, median flux density, and luminosity). 
Finally, there are two constraints on variability (how the source changes over time), also based on the EHT data: 230 GHz source-integrated variability and variability of the visibility amplitude. The unfolding story is neatly foreshadowed in \textcite[5]{EHT2022_SgrA_paper5}: ``[t]he selected constraints are heterogeneous, and it is not yet possible to combine them in a consistent, fully satisfactory way. Indeed, uncertainties in the data and the models are not well enough understood to make that possible.'' The EHT analysis reproduces the current measurements of the source's properties only as an ensemble: any single model of the source fails to display at least one of its properties. Put simply, no model is empirically adequate with respect to all of the observational constraints. 

We will now go through a high level summary of the models and their flaws, and then argue that this situation should be seen as involving epistemic opacity.

After introducing the constraints, \textcite{EHT2022_SgrA_paper5} consider two types of models: fiducial models and exploratory models. Roughly, the difference between these is that fiducial models share a number of core assumptions, corresponding to a kind of best guess about how to represent some important physical features; exploratory models modify at least one of these standard assumptions, allowing scientists to consider a wider variety of possibilities.

First are the fiducial models. These assume that the source is a Kerr black hole with a given mass and at a given distance, and use those parameters for setting up a variety of GRMHD simulations, from which images are produced through computation of the radiative transfer and ray tracing. These images can be compared with the Sgr A* observational constraints. It turns out that \textcite[30]{EHT2022_SgrA_paper5} ``[n]one of the fiducial models survive the full gauntlet of 11 constraints. If we set aside both variability constraints, however, there two fiducial models that pass the remaining nine constraints in all simulation pipelines (a few more survive in one model set but not the other).'' This is bad news: the fiducial models are not empirically adequate. But all fiducial models share some assumptions, including a particular prescription for the electron distribution function, a form of initial data (a magnetized torus), and posit that black hole spin and angular momentum of the torus either align or anti-align. If these assumptions of the fiducial models are to some extent flawed, they might introduce errors. 

The collaboration also considers a second class: exploratory models. These exploratory GRMHD models include: models with different prescriptions for the electron distribution function; models in which accretion is fed by stellar winds; and tilted disk models (in which black hole spin can be misaligned to angular momentum of the torus). Several types of these exploratory models are considered in \textcite[section 4.2]{EHT2022_SgrA_paper5}. Long story short, it turns out that six exploratory models initially pass all of the constraints, but the joy is short-lived: that success depends on the resolution with which they were computed. When these six exploratory models were re-analyzed with longer simulation runs, all of them failed at least one constraint.\footnote{It should be mentioned that the EHT does suggest a best-bet region of the MAD model space, under the assumption that the variability constraint is reduced by 30\%. This might be achievable due to the combination of \autocite[25]{EHT2022_SgrA_paper5} ``extended flux, viscosity, cooling, and numerical limitations'' (the latter include effects such as ``numerical inaccuracies in radiative transfer, truncation error in the GRMHD integrations (limited resolution), limited simulation duration, or misspecification of the adiabatic index'', although the collaboration does not find positive evidence that any of these numerical factors do in fact produce the searched for excess in variability.}

In summary, no single GRMHD model is able to account for all of the eleven observational constraints of Sgr A*; all the models are (to an extent) empirically inadequate for Sgr A*. Framing this as a test of the GRMHD models, \textit{no model passes the test}. This empirical inadequacy, by itself, need not signal any form of opacity of these models. However, we will now sketch an argument that the GRMHD models are indeed opaque in the sense that scientists cannot confidently determine which features of the model account for its success and failures.

We begin with the observation that the epistemic situation is one where GRMHD simulations and the properties of Sgr A* systems are being tested simultaneously. Validating these simulations (for these physical regimes) involves comparison to inferred properties of Sgr A*, while validation of the inferred properties relies on comparisons to GRMHD simulations. As is often the case in astrophysics, model validation faces a kind of holist challenge. Identifying what is going wrong in such a situation is difficult in part because of the diverse assumptions that are implicated in any test.\footnote{However, some of them can to some extent be disentangled. Some portion of uncertainty concerning the validity of various models can be reduced independently from observations of particular sources, or in tandem with them. One useful tool for this are code comparison projects, see \textcite{porth2019event} for the GRMHD code comparison project and \textcite{mizuno2021comparison} for a comparison of ion-to-electron temperature ratio prescriptions. Naturally, the Sgr A* analysis by EHT also involves independent lines of evidence in the four non-EHT observational constraints, and the uncertainties associated with these data are independent from the opacity of GRMHDs. For a philosophical discussion of these issues in the context of black hole astrophysics, see \textcite{Elder2023philastro,Elder-Doboszewski2025}. We thank an anonymous reviewer for insightful comments on these issues.}

In the situation just described, it is difficult to establish even whether the empirical inadequacy of the models are attributable to any of their inherent features. \textcite[p.14]{EHT2022_SgrA_paper5} summarizes the situation as follows: ``[i]t is likely that the models are physically incomplete. It is also possible, however, that one of the constraints is measured incorrectly, that one of the constraints is applied incorrectly, or that one of the constraints is poorly predicted for numerical reasons''. In other words, it is not clear what the failure of the modeling process should be attributed to: physical features of the models, the wrong value of a parameter, or flaws in the modeling process. Without being able to track how specific assumptions play out in simulation outputs (as would be the case for a fully transparent model), scientists cannot trace the relationship between specific assumptions and the empirical adequacy or inadequacy of their models. The scientists' ability to identify flaws in their models is thus limited. These GRMHD models are, then, opaque to its users in a sense that, to paraphrase \textcite[170]{Winsberg2018}, the features that account for the successes and failures of GRMHD models in simulating certain aspects of Sgr A* are not, in this case, able to be fully understood by the collaboration. This is in sharp contrast with the GRMHD models of M87*, where one can point to a clear class of the models reproducing properties of that source. 

While bothersome for the astrophysicists, this opacity might not be that surprising: GRMHD models are highly complex simulations; and, across sciences, complex simulations are prone to being opaque. Because the GRMHD models cannot recover all of the observational constraints, the opacity cannot be ignored and is seen as an important problem to be solved.\footnote{While the successes and failures of these models are not easy to pinpoint, many leads can be pursued. To get a sense of some of these, here are some ways the EHT is suggesting for the Sgr A* models \autocite[30]{EHT2022_SgrA_paper5}: ``The failure of nearly all fiducial models to match the light-curve variability is interesting. It may signal the presence of extended, slowly varying structure that is resolved out by EHT, or it may signal that future models need to incorporate collisionless effects (potentially modeled as viscosity and conductivity) or a more sophisticated treatment of electron thermodynamics including cooling. In addition, different initial geometries and polarities of the magnetic field could lead to more slowly varying structures. [...] If, when combined, these effects were to reduce M3 by 30\%, then many MAD models would be consistent with the data.''. However, currently this problem remains open.}

These limitations of the current scientific understanding of Sgr A* illustrate a way in which some GRMHD models are opaque to scientists. This opacity, naturally, is not due to use of ML methods --- no such methods were used in the 2022 analysis of Sgr A* by the EHT --- but rather due to the complexity of the models and the modeling process itself. 
It is not that they are uncertain whether an inference method succeeds in what it is doing; rather, they are uncertain about what should be modified in order to better account for the data, and even if the current data are correctly measured (and, so, if anything about the models themselves should be modified at all). So the adequacy of the source modeling remains open, the exact manner in which models are inadequate representations is unknown, and the relationship between modeling assumptions and empirical (in)adequacy is obscured. The opacity is thus quite worrisome. This form of opacity reveals that current understanding of the Sgr A* is limited, and GRMHD is a form of a black box that should be opened and looked into. But other black boxes, including ML black boxes, might not need to be.

At present, this opacity is a problem for interpretation of black hole images, but not for the production of the images themselves (which do not rely on assumptions from GRMHD). However, as we will see later (in sections \ref{sec:ML_BH_imaging} and \ref{sec:ourtake}), at least one prominent use of ML in black hole imaging relies heavily on GRMHD simulations, and consequently inherits the opacity and potential bias of these simulations.

\section{Machine Learning Methods in Astrophysics}\label{sec:ML_astro}

ML is used in a wide variety of applications in astrophysics. 
While (at time of writing) the EHT has not yet incorporated ML methods into its official publications on imaging of supermassive black holes, a number of possibilities have been explored, and an image using ML has been produced (i.e., published by researchers affiliated with the EHT, but not published by the collaboration). 
In this section we consider (in section \ref{sec:catalogues}) an illustrative example of how ML has been incorporated into a different astrophysical experiment. The use of AutoML to analyze Hubble data has similar motivations to those prompting consideration of ML for EHT imaging---in particular, achieving greater computational efficiency in response to a data analysis bottleneck. This example helps motivate the idea that one can judge the reliability of a ML method (AutoML) without opening the black box to see how it works. 
Then, in section \ref{sec:ML_BH_imaging}, we describe three (of many) proposed applications of ML to radio astronomy and black hole imaging---R2D2, $\alpha$-DPI, and PRIMO. We lay out how they work and the ways in which each may be considered opaque. 
Overall, sections \ref{sec:catalogues} and \ref{sec:ML_BH_imaging} motivate and provide background for our analysis of how to judge the reliability of ML methods in astrophysics in section \ref{sec:ourtake}.

\subsection{ML methods for large catalogs}\label{sec:catalogues}

ML is used by \textcite{kruk2022hubble} in the search for asteroid trails in the Hubble space telescope archive. This project combines citizen science with machine learning based on a DNN. Its broad aim was to produce a classifier that automatically works on new data as they are continuously added to the archive.

In the first year-long citizen science project phase, 11,482 volunteers identified 1,488 asteroid trails (on the set of images collected up to 24 April 2020). The results of their classification were then used as a training dataset for automated ML algorithm AutoML running on the Google Cloud Platform. AutoML was then applied to the full archive of images (collected up to 14 March 2021).\footnote{A similar methodology with a larger sample size has been used by \textcite{shimakawa2024galaxy} in search for ring galaxies in the data from SUBARU, an optical and infrared telescope located at Maunakea. There the construction of the training dataset involved circa 20,000 galaxies classified by humans; ML was then applied to around 700,000 galaxies in the catalog, and classified some 400,000 galaxies as spiral galaxies and some 30,000 as ring galaxies.} 
The main outcome was that 2487 new trails were found, out of which a total of 1701 have been confirmed by further analysis by the scientists themselves, 670 of which matched with known objects, mostly (95\%) in the Main Belt.

The question, then, is whether one should be concerned about the opacity of AutoML. It certainly might be interesting to know how and why AutoML made the decisions it actually did, and it might be an interesting black box to pry. Furthermore, if it turned out the method was prone to classification errors then the opacity might be a barrier to identifying and remedying the source of error (similarly to the Sgr A* example above). But, crucially, the inaccessibility of the inner workings of the AutoML does not seem to undermine judgments about its reliability as a classifier. This reliability judgment is founded on expert corroboration of its outputs, in addition to knowledge of the way that the DNN was trained (from classifications by citizen scientists). Furthermore, this case does not suffer from the worry of ``link uncertainty'' discussed by \textcite{sullivan2020understanding} (see section \ref{sec:opacity}) because there are connections between certain image features and the presence of asteroids that have already been empirically established---this background knowledge underpins the application of AutoML. 
The opacity or transparency of the classifier also seems to be irrelevant to the broader scientific goals here: utilizing the full Hubble archive and more ambitiously, of handling the 10+ Petabytes of forthcoming Euclid data. 

In effect, AutoML provides an example of the use of an opaque method in astronomy which does not seem to raise further epistemic concerns about its reliability. Three features of this example are worth noting. First, it involves a previously explored domain, where the performance of the ML model is well-benchmarked against an accepted standard. Second, other means of going through the (full) data are available in principle, although the use of ML to speed this up is what makes the task tractable in practice. Third, there is further human control over the results in the form of experts validating the outcomes of the classifications made by AutoML. This amounts to the availability of independent checks on the reliability of the method at each stage---in assessing the training data, validating the results of training, and then evaluating the final outputs from the full dataset. 

Together, these three features motivate our conditions C1-C3 below (section \ref{sec:ourtake}), which describe conditions under which ML can be reliably implemented in previously-explored physical regimes, where there are established benchmarks for testing ML-based results. Trust in the results of AutoML comes down to trust in the independent methods available for validating its outputs. However, a question then remains about whether ML can be used to obtain reliable results at the bleeding edge of our empirical grasp, where we lack previously established benchmarks. Even in the case of AutoML, when the method is applied to new, previously-unclassified data, we note that the availability of an independent classification method (expert judgment) provides an in-principle basis for a robustness argument to validate these results.\footnote{Insofar as expert judgment is already considered trustworthy for new data, such a robustness argument might not be needed.} The robustness arguments adopted by the EHT in the context of imaging both M87* (see \textcite{Elder-Doboszewski2024}) and Sgr A* (see section \ref{sec:sgra}) provides a further template for how to establish the reliability of methods when going beyond previously-explored physical regimes, motivating our condition C4 below (section \ref{sec:ourtake}). But before turning to our analysis of the conditions under which ML can be reliably implemented in black hole astrophysics, we will describe some of the ML methods that have already been put forward for use in this context. 

\subsection{Machine Learning Methods for Black Hole Imaging}\label{sec:ML_BH_imaging}

In this section we survey three examples of ML methods developed for radio astronomy and in particular for black hole imaging. They are motivated by different problems in various domains, from image production to parameter estimation, and therefore their eventual applications in the inference process would occur at different points. For the reader who wishes to skip the more technical descriptions of these methods below, we begin with a high-level summary: The first two of these methods (R2D2 and $\alpha$-DPI) explicitly utilize DNNs, so can be seen as involving epistemic opacity, for reasons discussed in section \ref{sec:opacity}. These two methods are motivated by the difficulties with scaling, or in other words, by the computational complexity and expense of the existing implementations of current inference methods. ML provides a faster way of computing functions which could be computed by other means. The third example is the PRIMO algorithm, which utilizes a method called ``dictionary learning'' to produce high fidelity images from sparse data. In general, dictionary learning is not considered to be opaque. However, the training dataset for PRIMO is composed of GRMHD simulations, which implicates PRIMO in epistemic opacity, albeit opacity that is not directly related to its being a ML method (see section \ref{sec:sgra}). PRIMO has been advertised as a way of generating ``physically motivated inferences for the unobserved Fourier components'' \autocite[1]{medeiros2023primo-image} which allows to generate a much sharper (by a factor of 2) image.
These examples are non-exhaustive\footnote{For the Event Horizon Telescope data, one use of neural networks might be extracting some physical parameters from visibilities, skipping the image reconstruction stage, proposed in \textcite{lin2021vlbinet} and more recently in \textcite{protopapas2025parameter}; another approach to imaging based on deep learning from closure quantities (which bypass the need for calibration) is DIReCT presented in \textcite{lai2025deep} and its continuation GenDIReCT; another example is ongoing work on Kinetic NeRF(or Kine) which disposes of priors defined by humans (which are seen as prone to bias) in favor of regularizations obtained by parametrizations of images through a neural network. Further examples are the Zingularity framework for parameter inference with deep Bayesian artificial neural networks \autocite{Zingularity2025-1,Zingularity2025-2,Zingularity2025-3}; yet another approach to deep learning for M87* can be found in \textcite{tsui2025deeplearning-m87}. More generally, for radio astronomy, various DNN-based ML approaches have been developed, including convolutional neural networks \autocite{connor2022deep-polish-cnn, schmidt2022deep-cnn, dabbech2022first} and generative adversarial networks \autocite{geyer2023deep-gan}. All four of the latter methods have been developed with the explicit goal of providing superresolution, that is, to construct images with resolution below that of the angular resolution of the telescope. PRIMO, discussed below, is also superresolving, an issue we will revisit in our discussion in section \ref{sec:ourtake}.}, but they illustrate well the variety of contexts and purposes for which ML methods are being developed.

Our first example of ML in radio-astronomy is R2D2, or ``Residual-to-Residual DNN series for high-Dynamic-range imaging''; our discussion follows \textcite{aghabiglou2024r2d2} (which builds on previous similar algorithms, AIRI and uSARA in \textcite{dabbech2022first}). Although this particular denoiser is currently developed for the Very Large Array (i.e., for radio astronomy, but not VLBI specifically), a VLBI array like the EHT faces similar scalability problems relating to the processing of increasing volumes of noisy data, and this or a similar approach might be eventually applied to the EHT data. Image reconstruction from a large dataset relies on various methods to separate signal from noise in the data. For this, R2D2 builds on developments of ``plug-and-play'' (PnP) ML algorithms.\footnote{In comparisons to Bayesian optimization algorithms, such SARA, PnP algorithms can achieve slightly better imaging precision at reduced computational cost \autocite{Terris-etal2023}. However, like optimization algorithms, PnP methods are highly iterative, meaning that they don't address scalability problems associated with large data volumes. In contrast, data-driven end-to-end DNNs, which are trained to produce images directly from the data (e.g., \textcite{connor2022deep-polish-cnn}) address issues of computational efficiency, but are said to lack interpretability and generalizability \autocite[2]{aghabiglou2024r2d2}. So, if that tradeoff is indeed unavoidable, then R2D2 tries to strike some form of balance between these competing epistemic virtues.} PnP methods involve training a dedicated denoising DNN, which is then deployed along with an optimization algorithm (which computes, in this case, an image compatible with the constraints given by the data, and tends to scale linearly with the data). R2D2 uses a series of DNNs to address the issue of scalability. (Without going into details, \textcite{aghabiglou2024r2d2} propose a variety of DNN architectures and show that some of them can be effective with just a few iterations.) Essentially, R2D2 reconstructs images iteratively in a manner reminiscent of the CLEAN algorithm: at each stage it produces an updated image estimate, based on the previous round's image estimate and associated residuals as input. But, for some number $I$ of iterations, $I$ separate DNNs are used as successive stages. Formally, the $i$-th image equals the $i-1$ image + (output of the DNN$_{i}$ which is given $i-1$-residual and $i-1$ image); see Figure 1 of \textcite{aghabiglou2024r2d2} for an illustration.

R2D2 is advertised as addressing both scalability and issues of interpretability and generalizability. Alas, neither \textcite{dabbech2022first} nor \textcite{aghabiglou2024r2d2} discuss the issue of interpretability in detail. The first of these papers states that a predecessor of sorts to R2D2, AIRI, ``inherits the robustness and interpretability of optimization algorithms'', and note that a particular constraint on the denoiser's loss function, firm nonexpansiveness, must be satisfied to ``preserve algorithm convergence and the interpretability of its solution''. In \textcite{aghabiglou2024r2d2} it is merely mentioned that end-to-end DNNs (which produce images directly from the data), while ultra-fast, lose ``robustness (interpretability and generalizability)''.\footnote{Note that the use of the term ``robustness'' to describe this method is different from way that we use it to apply to convergence of results across methods---following \textcite{Elder-Doboszewski2024}, agreement across independent lines of evidence. For philosophical analysis of the ways that ``robustness'' is used in ML contexts, see \textcite{Freiesleben-Grote2023}.} Generalizability is addressed in two ways: first, by normalization procedures \autocite[4]{aghabiglou2024r2d2}; and, second, because the training and validation datasets ``consist of 20000 and 250 pairs of ground-truth images and associated dirty images'' of two types: medical and optical astronomy images, which is intended to increase confidence that the DNNs trained on these data will generalize to other data \autocite[7]{aghabiglou2024r2d2}. The network is interpretable in the sense that the deep network is ``unrolled'' into a number of iterations, allowing access to partial stages of the implementation of a learning algorithm and their parametrizations; see \textcite{monga2021unrolling} for an overview of this approach. Assessment of the extent to which deep unrolling approach to interpretability can be philosophically questioned is beyond the scope of our argumentation. At a high level,  something that could be said is that we are obtaining more than just a final image from the data: R2D2 provides a sequence of $I$ images, split into the $i$-th image and its residuals, allowing partial access to the stages of imaging procedure. While the functioning of any particular DNN$_{i}$ remains a black box, the overall process is less mysterious because stage inputs can be easily accessed, and overall transition from, e.g., smooth structures to finer details and finer structures can be traced throughout iterations. That, effectively, trades a single black box for $I$ of them placed in a series, diminishing the influence of any single one of them on the overall process.

Our second example is $\alpha$-DPI (``$\alpha$-deep probabilistic inference''), a deep learning Bayesian inference approach which, among others, can be used to infer black hole features \autocite{sun2022alpha}. $\alpha$-DPI aims to quantify uncertainty in Bayesian inference problems, where some observed measurements $y$ are given, and one tries to estimate the posterior probability of some probability distribution $x$, $P(x|y)$. There are two types of traditional methods used for this: sampling-based approaches (such as importance sampling and Markov Chain Monte Carlo, ``MCMC'' methods) and optimization-based inference approaches (such as variational inference, ``VI''). The former are accurate but often prohibitively slow (something to keep in mind for our discussion of opacity in PRIMO below) and do not scale well for high dimensional data, while the latter can produce overly simplified distribution estimates \autocite[1-2]{sun2022alpha}. In this context, \textcite{sun2022alpha} propose to combine them: for EHT-like data, first a variational inference is performed with a trained neural network, producing samples of geometric parameters. The second step fits the image samples to the telescope data, and evaluates whether the samples are a good fit to the data through importance sampling. \textcite{sun2022alpha} apply $\alpha$-DPI to both simulated and actual EHT data in order to estimate properties of the source emission (such as diameter of the observed crescent shape, or brightness of its emission) and suggest that the efficiency of this approach will be useful in solving future scaling problems that will arise with an expanded ngEHT array.

Our third and final example is the PRIMO (``Principal-component Interferometric Modeling'') method. \textcite{medeiros2023primo-detailed} offers a detailed overview of this approach and \textcite{psaltis2024primo-theoretical} provides a review of its theoretical basis, including a discussion of possible biases. PRIMO has been used to construct a sharper image of M87* \autocite{medeiros2023primo-image}, using the same 2017 dataset as the original 2019 EHT image \autocite{2019EHT_M87_paper1}. PRIMO differs from the above two ML methods: most importantly for us, its current incarnation does not rely on a DNN, which reduces PRIMO's opacity at the level of the realization of the algorithm, in line with \textcite{Creel2020} distinctions. But it is a machine learning method, and an image created with it has been publicized as an example of the possible benefits of artificial intelligence. PRIMO relies on physical priors about what the image could be like, and as such, although this ML approach is less prone to the charge of opacity than those relying on DNNs, it has to address the issue of potential bias due to its reliance on opaque simulations.

The imaging algorithms the EHT used in 2019 and 2022 try to remain theory-neutral. Nevertheless, some assumptions are needed to overcome the ill-posed inverse problem of producing images from sparse data; these algorithms attempt to fill in the areas of the domain for which the data are lacking by imposing various general but theory-neutral image features: smoothness, positive flux values, minimization of image gradient, compactness of the source, and so on.\footnote{Whether these priors are truly theory-neutral is debatable. For example, as \textcite{psaltis2024primo-theoretical} point out following \textcite{psaltis2015general}, a lack of sharp edges might seem a reasonable feature for images to have. However, a theoretical expectation is that, at high resolution, the emission of the black hole shadow does drop sharply on the inside part of the image.} PRIMO, instead, is trained on principal components or eigenimages whose superpositions produce model images, and uses those components in filling in the gaps in the sparse coverage of EHT data. This amounts to sparse dictionary learning, in which principal components function as atoms and images generated with them as the dictionary. PRIMO tries to obtain an optimal image for given components by minimizing a loss function between components of the model image and the components given by the data.

Physical assumptions enter with the choice to use GRMHD simulations to generate the training dataset. This training set (described in details in section 2.1 of \textcite{medeiros2023primo-detailed}) consists of a library of 30 x 1024 (30 720 total) snapshots of GRMHD simulations; in effect, an application of PRIMO assumes the validity of that set as a possibility space for what the source can be like. The observed components of the data are then fitted to simulations with similar values of their components. This is done through a dictionary-learning phase implemented with MCMC.\footnote{Dictionary learning could also be implemented with a DNN, which would increase PRIMO's opacity at the level of the realization of the algorithm, following the taxonomy given by \textcite{Creel2020} (see also section \ref{sec:opacity}).} Because ``[w]ithin a given component, structural information on fine angular scales is correlated with that on broader scales'' \autocite[6]{medeiros2023primo-detailed}, if the broader scale is set to the resolution of the array, PRIMO nevertheless entails information about structures below it, providing superresolution.

Further physical assumptions are made about the physical parameters of the simulations. Examples of these include the choice of the particular value of the black hole's spin (e.g., $a = 0.9$), assumed to be aligned with the black hole jet.\footnote{The dimensionless spin parameter $a$ quantifies the rotation of the black hole as a fraction of the maximum possible value (within the Kerr family of black hole solutions). Apart from physical effects (such as the presence of an ergosphere) and astrophysical significance (such as the efficiency of the conversion of accreting matter into electromagnetic radiation), this parameter also has a qualitative consequence for the spacetime geometry. Above the value 1 the event horizon would no longer be present and the central singularity would become visible to external observers.} For this particular assumption, it turns out that an algorithm trained on this basis can reconstruct simulated images corresponding to different spin values, making outcomes of PRIMO only weakly dependent on that particular choice for the parameter. Other choices---e.g., concerning ion-to-electron temperature ratio, electron density scales, and MAD and SANE accretion models---are also made, each spanning some value range. The mass range for the simulations in the training dataset for the M87* image with PRIMO was set to scalings of the black hole mass ($M = 6.5 \times 10^{9} M_{\odot}$); these scalings affect the image's size and the radiative transfer calculation. This mass value, in turn, was fixed by the converging value of previous stellar dynamics measurement and the 2019 EHT estimate of the black hole mass, based on the same 2017 dataset (diverging from the value of gas dynamical measurements of the same quantity). This iterative re-analysis of the 2017 data thus assumes, to some extent, the validity of the 2019 image. In effect, in making an inference about the unobserved components PRIMO indeed uses strong physical priors in generating the training image dataset used to match data with the models.

\section{Establishing Reliability: When to Trust the Results of Machine Learning in Astrophysics}\label{sec:ourtake}

We have now set the stage with three main threads of analysis: first, an overview of opacity; second, an account of how it occurs in GRMHD simulations of black hole spacetimes; and third, a description of three uses of ML in black hole imaging. Now, we will draw these threads together in order to make the case that the opacity of an inference method does not (always) diminish trust in its outcome. In other words, we think that opaque methods can play an important role in establishing new scientific results and even in gaining understanding about physical systems. Opacity does not always need to be reduced. We will first articulate what we take to be conditions under which opaque methods may be reliable and informative, and then extend the discussion of these conditions to the examples from black hole imaging discussed in section \ref{sec:ML_BH_imaging}.

Our position on ML is cautious and conservative, in that we would consider a detection claim based primarily on ML to be rather unreliable in the absence of either independent evidence or an argument for its reliability. We take this stance because even without compounding the issue with opacity and training dataset biases, overconfidence in modeling approaches has led to detection claims which, while bold in a Popperian sense, turned out to be not reproducible. Astronomical examples of that include the BICEP2 claim concerning detection of primordial gravitational wave signatures \autocite{BICEP2014}, and the claimed detection of the photon ring based on EHT data \autocite{Broderick-etal2022,Tiede-etal2022,Lockhart-Gralla2022}.\footnote{See \textcite{Elder202Xphoton_ring} for philosophical discussion of this case.}

Nevertheless, the example of the use of AutoML for a large catalog in section \ref{sec:catalogues} suggests that, in some astronomical contexts, a careful use of an opaque ML method does not need to be associated with allegations of either bias, unreliability, or lack of understanding associated with the outcome. In such cases, the use of ML may be part of a reliable method for learning about the target system. The case of AutoML also illustrates how the broader experimental context matters when evaluating these issues. We first discuss two minimal conditions, and then two additional ones, aimed at previously explored domains and at novel domains.

\subsection{A reliability assessment framework}

Comparing such cases to existing, non-ML examples of EHT imaging methods suggests that the following minimal features have to be satisfied (as necessary conditions) when judging the reliability of an inference relying on an algorithm (including, but not limited to, ML):

\begin{itemize}
    \item [\textbf{(C1)}] \textbf{Training data bias mitigation.} The algorithm's training dataset should be large and diverse enough to allow for a wide range of physically reasonable possibilities.
    \item [\textbf{(C2)}] \textbf{Bias tracking.} Bias that may be introduced through choices made in the implementation of the algorithm should be tracked.
\end{itemize}

(C1) states that care needs to be taken to ensure an appropriate training dataset. This means that the training set needs to include data corresponding to a wide range of physical scenarios that might be present in the target system. For example, training a DNN based on a subset of GRMHD simulations corresponding to our current best estimates for M87* might be overly narrow and bias results towards reproducing expected outcomes (i.e., a black hole image that looks like the famous 2019 one, with values for physical parameters conforming to prior estimates). In the case of the non-ML imaging algorithms used to image M87*, the training data for the regularized maximum likelihood algorithms (eht-imaging and SMILI) included not only a full library of GRMHD simulations but also a number of geometric models (ring, crescent, disk, double Gaussian) included to deliberately expand the range of possibilities being accounted for.

One instructive failure of (C1) is a failure of transferability. This is neatly illustrated by an example from the use of machine learning in cosmology. \textcite{villaescusa2022cosmology} worked with 2000 simulated cosmological models and around 1 million galaxies. A trained ML model predicted the overall matter density with up to 10\% accuracy on the basis of a density of a single galaxy, a remarkable success. But ``if the models are trained on galaxies from the IllustrisTNG simulations, they cannot infer the value of $\Omega_{m}$ from galaxies of the SIMBA simulations, and vice versa'' \autocite[13]{villaescusa2022cosmology}.\footnote{Here $\Omega_{m}$ is a parameter representing the matter density in the Friedman equations in cosmology. Roughly, this quantifies the energy density of the universe due to matter (including both ordinary baryonic matter and dark matter).} This may be due to intrinsic differences between these simulations (but due to the opacity of the network, this cannot be easily ascertained; in terms of w-opacity \autocite{Boge2022}, what the network learned is unclear).\footnote{Note that \textcite{villaescusa2022cosmology} call this a failure of ``robustness'' rather than ``transferability''; but we prefer, in accordance with \textcite{Elder-Doboszewski2024} and other extensive literature summarized therein, to use that term in a more strict way to denote convergence of independent lines of evidence.} One of the aims of (C1) is to minimize the chance that this kind of an unfortunate situation occurs.

(C2) postulates that the ways in which new methods might be biased is something to be explicitly studied, controlled, and tracked throughout applications. Two key sources of bias are worth particular attention: first, theoretical bias introduced by the assumption that current theory (e.g.,  general relativity and accretion physics, as assumed by GRMHD models) provides an adequate description of the target system; and second, that (GRMHD) models used in the dataset are a representative sampling of the possibility space. This is a diachronic condition, which, for ML, might be of increasing importance as methods with various forms of potential biases are developed and implemented. Importantly, in section \ref{sec:sgra} we argued that, for some important sources, whatever the ground truth model really is, it is likely not quite any one of the currently available ones, and therefore training relying on GRHMD simulations might already be biased.

These two conditions are not sufficient. The further following condition captures the context-dependency of the reliability of inferences with opaque methods.

\begin{itemize}
    \item [\textbf{(C3)}] \textbf{Benchmarking in explored domains.} The algorithm should be tested in a physical domain to which we have prior empirical access before being used to make inferences about it.
\end{itemize}

(C3) says that a new algorithm should first be applied in domains that have already been explored by other means. This criterion is motivated by a very much open possibility that currently-accepted theories may not provide accurate descriptions of previously unexplored domains. Because of this, additional empirical evidence would be needed in situations where (C3) does not hold. For example, the LIGO-Virgo observations of black holes provided the first empirical access to the dynamic strong field regime of general relativity. In advance, there was no guarantee that general relativistic models would be adequate in this regime. Without (C3), validating ML methods applied to new domains of applicability would be challenging, if not impossible, because there would be no trustworthy training data, and no reliable benchmarks to test its results against.\footnote{For a recent philosophical discussion of benchmarking in machine learning, see \textcite{freiesleben2025benchmarking}.}

Of course, what is meant by a new domain or regime may be a subtle matter. For the EHT itself, there are at least two ways to think about the relevant regimes. With respect to inferences about the physical properties of the black hole region, the relevant domain is determined by the physical properties of the system being observed: a relativistic spacetime with strong gravitational effects along with the physical conditions associated with an accretion disk. But for the purposes of imaging itself, the relevant domain might rather be determined by the angular resolution of the array and its application to sources at the limits of that resolution---including dynamic sources that change on timescales relevant for M87* and Sgr A*.

In the context of the ngEHT, there is a clear scope to use opaque ML methods for monitoring campaigns, where the same targets (M87* and Sgr A*) might be observed at a comparable resolution to that of the initial campaigns in order to determine which features are stable and which features change over time. In this case, (C3) would clearly be met. In that domain, if needed, a further check could also be performed on a case by case basis through comparing ML results to those produced by the methods used for initial imaging of these targets. In monitoring campaigns many surprises (in the sense of new phenomena or unexpected changes in the target system) are possible. In such cases, comparing the results of ML methods to other methods would be important, in order to build confidence that these surprises were not an artifact of the method (something that is not easy to identify for opaque methods when applied in isolation).

When (C1)-(C3) are all satisfied, errors and biases introduced through the use of ML methods could be kept in check by comparing whether they continue to reliably perform the computational tasks asked of them, even if these methods themselves are completely opaque. 

So far, this is not a surprising result. Few would object to the importance of (C1) and (C2), while (C3) amounts to a conservative condition: we can trust an opaque ML method in domains where it can be tested against well-established results. However, this result is already of importance for the purposes of black hole astrophysics. Recall that we started by pointing out a significant bottleneck problem: the EHT cannot analyze data fast enough with existing methods. For the purpose of monitoring campaigns---observations that amount to more data about previously observed systems at previously-observed resolution---the ability to confidently implement ML methods could be crucial to speeding up analysis to keep pace with data collection. As well as decreasing the wait on any new scientific results, solving this problem could be crucial in making the case for the value of the EHT and its future expansions in the context of competitive funding applications.    

However, we also want to consider the role of ML methods in situations beyond those where (C3) holds. Could ML ever be trusted in an extended, novel, domain? In particular, for the EHT, what role could ML play in producing new higher-resolution images or in making other inferences about supermassive black holes that go beyond what has already been done?

The following additional condition addresses these questions:
\begin{itemize}
    \item [(\textbf{C4})] \textbf{Robustness.} The results of a ML method should converge with results based on independent methods or lines of evidence.
\end{itemize}
This says that ML methods can be used as part of a robustness-type inferential strategy in which confidence in the outcome arises from the convergence of multiple independent lines of evidence. This is the strategy already used by the EHT for imaging and inferences (e.g., parameter estimation) with standard, non-ML algorithms.\footnote{From the conceptual point of view, this general strategy is discussed at length in \textcite{Elder-Doboszewski2024}, in conversation with broader discussions of robustness in modeling and experimental contexts (e.g., \textcite{Staley2004robustness,Staley2020,Dethier2022,schupbach2018robustness,ritsonstaley2021uncertainty,Parker2011,parker2020model,Winsberg2018,karaca2020two,bogerobustnes,Gueguen2020,Cartwright1991}.} For the purpose of imaging, this requires that ML methods be used alongside other accepted methods (such as industry-standard CLEAN) and that its results be trusted insofar as there is agreement across methods, and their independence can be demonstrated. It may even be viable to use multiple ML methods as the basis for such a robustness argument, provided that these methods are demonstrably independent. The full robustness argument does not need to be provided every time; once agreement has been demonstrated, it may be sufficient simply to have an option to check against other methods on a case-by-case basis, rather than replicating the process with different imaging methods each time.

\subsection{Assessing the reliability of black boxes in black hole imaging}

Let us now consider conditions (C1)-(C4) for the methods discussed in section \ref{sec:ML_BH_imaging}.

As for the \textbf{training data bias mitigation} of (C1), having a diverse enough training dataset, or, to borrow a phrase from \textcite[2]{aghabiglou2024r2d2}, diversified ground truth database, is something everyone seems to agree on in principle. \textcite[4]{medeiros2023primo-detailed} use similar terminology when they state that ``very large number of realistic simulated black-hole images'' and argue that ``[i]t is this broad range of possibilities encoded in the eigenimages learned by the algorithm together with the presence of high-quality data [...] that allow for PRIMO to recreate a robust image that depends only marginally on the training set used.'' \autocite[10]{medeiros2023primo-detailed}. However, in practice, the actual composition of the training dataset varies widely, and there seems to be little consensus as to what it should consist of. Two of the methods of section \ref{sec:ML_BH_imaging} already show this. $\alpha$-DPI fits geometric models to the data, and parametrizes black hole images as sums of asymmetric rings and elliptical Gaussians. In contrast, the PRIMO training dataset consists of GRMHD snapshots of various morphologies. These dissimilar training datasets render direct comparison difficult. Given that these methods are presently used for different purposes, comparing the two may be unproblematic. Nonetheless, addressing such concerns might become more challenging for future comparisons between different ML methods for black hole imaging. However, common benchmarks from the simulated and real EHT data (discussed below in the context of (C2)) might remedy this concern to some extent. In the meantime, perhaps these training datasets are best seen as simply exploratory, and the collaboration will develop a shared set of benchmarks over time.

The application of ML to black hole imaging relies on a training dataset, which is where (C2) \textbf{bias tracking} becomes particularly important. For physics-informed ML, such as PRIMO, this training dataset builds in assumptions about the source based on theoretical models. At present, the best models available come from GRMHD simulations, so PRIMO's training data is based on these. However, GRMHD simulations are known to have some empirical inadequacies: as discussed in section \ref{sec:sgra}, GRMHD simulations cannot not provide a fully satisfying model of one important EHT target, Sgr A*. No single GRMHD simulation is able to recover the values of all of the observed parameters; they only do so as an ensemble of partially correct models. This does not mean that training some algorithms on the results of GRMHD simulations is a bad thing to do. It is reasonable to take the best theory available and use it in an analysis. However, insofar as the training dataset for PRIMO reflects these inadequacies of GRMHD models, the imaging method trained on them might become biased during the training process, especially for sources --- such as Sgr A* --- in which the true model is outside the distribution used in training. This may result in less trustworthy outputs. 

In the context of M87*, for which the opacity of GRMHD is less worrisome, PRIMO has been tested in a few different ways: by recovering various morphologies of the emission \autocite[4]{psaltis2024primo-theoretical}; by varying the assumptions behind the GRMHD simulations (this retains the general form of the background assumption, but relaxes the particular physical values taken) \autocite[9]{psaltis2024primo-theoretical}; and by reconstructing images of black holes when the algorithm is trained on principal components obtained not from GRMHD, but from compact red-noise structures \autocite{hallur2022red}. This suggests that PRIMO could be trained on models very different from the data and nevertheless deliver the accurate outputs. However, this ends up requiring more principal components than the reconstruction from GRMHD ones, so it seems that in PRIMO imaging of M87* the stronger physical priors are traded for this increase in efficiency.

(C2) is a diachronic condition, and given relative nascency of ML methods for black hole imaging, it is overall hard to evaluate it at this stage. One natural and emerging form of bias tests is through the use of synthetic and real EHT data analysis results as benchmarks. For example, \textcite{sun2022alpha} compare the estimates of $\alpha$-DPI to existing parameter estimates (they consider THEMIS and dynesty as two benchmark pipelines for the EHT feature extraction). More generally, it seems that, presently, the only shared ML benchmark is recovery of the outcomes of the previous EHT analyses.

Worries about the opacity of GRMHD arise for some sources (Sgr A*) but not necessarily for others (M87*, in which there is a clear class of models passing all the constraints). One could therefore see the downstream opacity as not particularly worrying for a source such as M87*. This ties well with the \textcite{parker2020model} adequacy-for-purpose framework: training dataset fit for one source does not need to be fit for another one.
\footnote{We are grateful to an anonymous reviewer for pressing us on this point.}

As for (C3), \textbf{restriction to explored domains}, for each of the ML methods we have discussed--- possible extensions of R2D2, $\alpha$-DPI, and PRIMO---the goal is do a task that the EHT has performed before for both M87* and Sgr A*. As such, it might look like they are all being applied within a previously explored domain, whether we cash this out in terms of the physical properties of the system or the angular resolution of the imaging. However, PRIMO also can achieve a higher resolution than the previously-reported EHT images by making use of GRMHD simulations to learn ``the correlations between the low-frequency and high-frequency structure'' \autocite[3]{medeiros2023primo-image}. PRIMO is thus a method providing superresolution, i.e., it gives information about structure going beyond the resolution of the array. Learning algorithms trained on sufficiently high resolution simulations can do that, because they correlate low resolution or general features with their high resolution underlying components (which might be visible only with the resolution going above that of the actual array).

In general, practitioners are wary of superresolution. Proponents of PRIMO provide what they consider the reliability limit of PRIMO (which they consider to be the sum of the largest baseline length and the baseline length that resolves the image), and \textcite[10]{psaltis2024primo-theoretical} note that ``[s]tructures with angular sizes smaller than this limit cannot be inferred from the data, whether one uses machine learning algorithms or not, and will necessarily require generating information that is not constrained by observations''. Accordingly, for M87*, \textcite{medeiros2023primo-image} provide both the superresolved image and a blurred image set to the nominal resolution of the array. So, while they do use PRIMO to go beyond previously explored domains, the paper also provides a more conservative image that accords with our (C3). While superresolving features of ML methods might have some value for exploratory experimentation, for instance as bets concerning future observations, or as prediction extraction from various models, conclusions drawn from superresolving features should not be considered reliable. This is because nothing like a (C4) can be established in such contexts, so any bias and error remain unchecked.

Finally, the last component of our account is (C4) \textbf{robustness}. For (C4) the context of an inference being made is crucial. The first two methods we discussed in section \ref{sec:ML_BH_imaging} could be employed as part of a robustness-style argument because they both present computationally efficient alternatives to existing non-ML methods used by the EHT. A variant of R2D2 adapted to VLBI could be deployed alongside existing imaging algorithms that use CLEAN or RML methods. Similarly, $\alpha$-DPI is performing a task already being undertaken with non-ML methods.

In contrast, PRIMO would be harder to fit in as a part of a robustness-type argument, because it relies on previous results in order to set the strong physical priors. As we said in section \ref{sec:ML_BH_imaging}, PRIMO builds on past results (such as mass estimates and validity of GRMHD models) and so, to some extent, assumes the validity of these results while attempting to infer more from the same data. This does not preclude a robustness-style argument for the reliability of PRIMO, but it does make independence more difficult to establish.
For example, if the physical priors used by PRIMO rely to some extent on results from the CLEAN and RML methods, then, with all of these methods applied to the same dataset, one cannot straightforwardly compare or average the results of PRIMO with those of CLEAN and RML to establish the robustness of the results. 

\section{Conclusion}\label{sec:conc}

Let's revisit the abstracted decision of section \ref{sec:opacity}: should the (next generation) EHT Collaboration employ ML methods for black hole imaging? And if so, should we demand that these methods be either explainable or interpretable? 

In this paper we have argued that opaque machine learning methods can play an important role in the future of black hole imaging; their epistemic opacity does not automatically render them unreliable. Indeed, the most troubling source of opacity among the cases we considered had little to do with the complexity of ML methods themselves.  
 
In section \ref{sec:ourtake}, we articulated the conditions under which implementations of opaque methods in astrophysics can be considered reliable. We considered two main cases.

First, one can trust the results of black boxes (ML or otherwise) when: (C1) the algorithm is trained on a sufficiently diverse dataset; (C2) potential sources of bias are tracked; and (C3) the algorithm is benchmarked and applied in previously explored domains. When these conditions are met, the reliability of an opaque algorithm can be established, obviating the need for explainable or interpretable models. This is important because it licenses the use of ML methods for monitoring EHT sources (M87* and Sgr A*) as new data accumulates. 
The use of a computationally efficient, reliable ML has the potential to solve the problem of data analysis bottlenecks, allowing scientific results to keep pace with data collection. 

Second, black boxes (again, ML or otherwise) can also play a part in generating reliable inferences in new physical regimes, such as future higher-resolution regimes for the EHT.    
In this case, using a ML method to push into higher-resolution regimes by itself is unreliable. For example, PRIMO's higher resolution images cannot be trusted, in part because they rely on GRMHD simulations and inherit potential problems of opacity and bias from these simulations (at least for Sgr A*). 
However, if (C4) holds and a particular ML method is used alongside other, independent methods, then a robustness argument may be used to justify confidence in convergent results. In this way, existing EHT methodology provides guidance for how opaque ML methods might be incorporated into scientific inferences without undermining their reliability.
For PRIMO however, such a robustness argument may not be readily available because the method uses physical priors derived using other imaging methods.

Our analysis resonates well with process reliabilism (or, more specifically, computational reliabilism), with context-dependent empirical caveats. In particular, we have argued that Sgr A* provides two closely related cautionary stories: one about the limitations of the GRMHD simulations themselves and another about the limitations of ML models that rely on GRMHD in their training. Sgr A* illustrates that there is a form of opacity already present in black hole imaging, and it is worrisome in ways that the opacity of ML methods might not be. As a consequence of this and our analysis of reliability in section \ref{sec:ourtake}, one of the most pressing reasons for worrying about the reliability of ML methods has little to do with ML \textit{per se}, but rather with reliance on GRMHD simulations during training. 

Overall, the key upshots of our paper are as follows. 
First, the EHT can implement ML methods for both monitoring campaigns (if C1-C3 are met) and for future observations with expanded arrays and/or at higher resolution (if C4 also holds). There seems to be little reason to push for the ML methods to be explainable or interpretable, as long as their reliability can be so-established.
Second, philosophers of science can take the EHT methods for establishing the reliability of their imaging pipelines as an exemplar for incorporating opaque ML. Given the challenging epistemic situation faced by the EHT, the ability to establish trust in black boxes in this case should make us optimistic about finding similar (if context-dependent) arguments for the reliability of opaque methods in a range of other scientific contexts.

\section*{Acknowledgments}

We are grateful to Martin King, Emanuele Ratti, audiences at: the DPG meeting in Berlin; JEHA-III; the Kraków and Oviedo and Friends workshop; the Boston Network for History and Philosophy of Physics; and the University of Otago; participants in: the Lichtenberg group WIP meeting; the Epistemology of Modern Physics reading group; the Rotman Philosophy of Physics Reading Group; the Philosophy of Machine Learning MCMP reading group, and the ngEHT Algorithms and Inference science working group for discussions of these and related ideas.

Early stages of this work were made possible by funding from the Lichtenberg Grant for Philosophy and History of Physics by the Volkswagen Foundation. JD's work on this project has been supported by the Polish National Agency for Academic Exchange (NAWA) under the Polskie Powroty program, grant number BPN/PPO/2024/1/00018. This project/publication is funded in part by the Gordon and Betty Moore Foundation (Grant \#13526). It was also made possible through the support of a grant from the John Templeton Foundation (Grant \#63445). The opinions expressed in this publication are those of the author(s) and do not necessarily reflect the views of these Foundations.

\newpage
\printbibliography
\end{document}